\documentclass[conference]{IEEEtran}
\usepackage{svg}
\usepackage{amsmath}
\usepackage[xindy]{glossaries}
\usepackage{graphicx}
\usepackage{subcaption}
\usepackage{epstopdf}
\usepackage{amssymb}
\usepackage{amsthm}
\usepackage{cite}
\usepackage{enumerate} 
\usepackage{tikz}
\usepackage{fancyhdr}
\usepackage{lettrine}
\usetikzlibrary{patterns}
\usetikzlibrary{shapes,arrows}
\usepackage{verbatim}

\usetikzlibrary{positioning}
\usetikzlibrary{shapes.geometric}
\usetikzlibrary{shapes.misc}
\usepackage{booktabs}
\usepackage{algorithm}
\usepackage[noend]{algpseudocode}
 \usepackage{siunitx}

\usepackage[font=small,labelfont=]{caption}

\newacronym{ai}{AI}{artificial intelligence}
\newacronym{aoa}{AOA}{angle-of-arrival}
\newacronym{aod}{AOD}{angle-of-departure}
\newacronym{ap}{AP}{access point}
\newacronym{asic}{ASIC}{application specific integrated circuit}
\newacronym{awgn}{AWGN}{additive white Gaussian noise}
\newacronym{blue}{BLUE}{best linear unbiased estimator}
\newacronym{ber}{BER}{bit error rate}
\newacronym{ce}{CE}{channel estimation}
\newacronym{cw}{CW}{continuous wave}
\newacronym{cdf}{CDF}{cumulative distribution function}
\newacronym{ced}{CED}{cumulative energy density}
\newacronym{clk}{CLK}{clock}
\newacronym{cpu}{CPU}{central processing unit}
\newacronym{csi}{CSI}{channel state information}
\newacronym{csp}{CSP}{contact service point}
\newacronym{crlb}{CRLB}{Cram\'er-Rao lower bound}
\newacronym{cu}{CU}{central unit}
\newacronym{dc}{DC}{direct current}
\newacronym{dl}{DL}{downlink}
\newacronym{dmimo}{D-MIMO}{distributed multiple-input multiple-output}
\newacronym{ds}{ODS}{digital scheme}
\newacronym{dlt}{DLT}{Distributed Ledger Technology}
\newacronym{dp}{DP}{differencial privacy}
\newacronym{dram}{DRAM}{dynamic random-access memory}
\newacronym{du}{DU}{distributed unit}
\newacronym{ecc}{ECC}{elliptic curve cryptography}
\newacronym{ecdlp}{ECDLP}{elliptic curve discrete logarithm problem}
\newacronym{ecsp}{ECSP}{edge computing service point}
\newacronym{eirp}{EIRP}{equivalent isotropically radiated power}
\newacronym{em}{EM}{electromagnetic}
\newacronym{embb}{EMBB}{enhanced mobile broadband}
\newacronym{en}{EN}{energy neutral}
\newacronym{e2e}{E2E}{End-to-End}
\newacronym{fa}{FA}{federation anchor}
\newacronym{fdd}{FDD}{frequency division duplexing}
\newacronym{fhss}{FHSS}{frequency hopping spread spectrum}
\newacronym{fl}{FL}{federated learning}
\newacronym{gdpr}{GDPR}{general data protection regulation}
\newacronym{ia}{IA}{initial access}
\newacronym{ic}{IC}{integrated circuit}
\newacronym{ioe}{IoE}{internet of everything}
\newacronym{iot}{IoT}{internet of things}
\newacronym{id}{ID}{information decoding}
\newacronym{kpi}{KPI}{key performance indicator}
\newacronym{lca}{LCA}{life cycle assessment}
\newacronym{ls}{LS}{least squares}
\newacronym{lis}{LIS}{large intelligent surface}
\newacronym{liion}{Li-ion}{lithium-ion}
\newacronym{llr}{LLR}{log-likelihood ratio}
\newacronym{los}{LoS}{line-of-sight}
\newacronym{lmo}{LMO}{lithium ion manganese oxide}
\newacronym{lsfc}{LSFC}{large scale fading coefficient}
\newacronym{map}{MAP}{maximum a posteriori}
\newacronym{mf}{MF}{matched filter}
\newacronym{ml}{ML}{maximum-likelihood}
\newacronym{mle}{MLE}{maximum likelihood estimator}
\newacronym{mvu}{MVU}{minimum variance unbiased}
\newacronym{mse}{MSE}{mean-square error}
\newacronym{mmse}{MMSE}{minimum mean square error}
\newacronym{lmmse}{LMMSE}{linear minimum mean-square error}
\newacronym{ldpc}{LDPC}{low-density parity-check}
\newacronym{mmwave}{mmWave}{millimeter wave}
\newacronym{mimo}{MIMO}{multiple-input multiple-output}
\newacronym{miso}{MISO}{multiple-input single-output}
\newacronym{mpc}{MPC}{multipath component}
\newacronym{mrc}{MRC}{maximum ratio combining}
\newacronym{mrt}{MRT}{maximum ratio transmission}
\newacronym{nmse}{NMSE}{normalized mean square error}
\newacronym{nods}{NODS}{non-orthogonal digital scheme}
\newacronym{ofdm}{OFDM}{orthogonal frequency-division multiplexing}
\newacronym{ota}{OTA}{over-the-air}
\newacronym{ods}{ODS}{orthogonal digital scheme}
\newacronym{nb}{NB}{narrowband}
\newacronym{nr}{NR}{New Radio}
\newacronym{pa}{PA}{power amplifier}
\newacronym{pdp}{PDP}{power delay profile}
\newacronym{per}{PER}{packet error rate}
\newacronym{pet}{PET}{privacy enhancing technology}
\newacronym{pki}{PKI}{public key infrastucture}
\newacronym{pl}{PL}{path loss}
\newacronym{pc}{PC}{pilot count}
\newacronym{qrrls}{QR-RLS}{QR decomposition based recursive least squares}
\newacronym{qam}{QAM}{quadrature amplitude modulation}
\newacronym{rb}{RB}{resource block}
\newacronym{re}{RE}{radio element}
\newacronym{rf}{RF}{radio frequency}
\newacronym{rfid}{RFID}{radio frequency identification}
\newacronym{ris}{RIS}{reflective intelligent surface}
\newacronym{rms}{RMS}{root mean square}
\newacronym{rls}{RLS}{recursive least squares}
\newacronym{rss}{RSS}{received signal strength}
\newacronym{rssi}{RSSI}{received signal strength indicator}
\newacronym{ru}{RU}{radio-unit}
\newacronym{rw}{RW}{RadioWeaves}
\newacronym{sa}{SA}{synchronization anchor}
\newacronym{sdg}{SDG}{Sustainable Development Goal}
\newacronym{sdn}{SDN}{software-defined network}
\newacronym{se}{SE}{spectral efficiency}
\newacronym{ser}{SER}{symbol error rate}
\newacronym{sinr}{SINR}{signal-to-interference-plus-noise ratio}
\newacronym{siso}{SISO}{single-input single-output}
\newacronym{slam}{SLAM}{simultaneous localization and mapping}
\newacronym{snr}{SNR}{signal-to-noise ratio}
\newacronym{srls}{SRLS}{standard recursive least squares}
\newacronym{svd}{SVD}{singular value decomposition}
\newacronym{sp}{SP}{service point}
\newacronym{sram}{SRAM}{static random-access memory}
\newacronym{tdd}{TDD}{time division duplexing}
\newacronym{tdoa}{TDOA}{time-difference-of-arrival}
\newacronym{toa}{TOA}{time-of-arrival}
\newacronym{tx}{TX}{transmitter}
\newacronym{rx}{RX}{receiver}
\newacronym[longplural={user-equipment}]{ue}{UE}{user-equipment}
\newacronym{uatf}{U-at-F}{use-and-then-forget}
\newacronym{uhf}{UHF}{ultra high frequency}
\newacronym{ula}{ULA}{uniform linear array}
\newacronym{upa}{UPA}{uniform planar array}
\newacronym{ul}{UL}{uplink}
\newacronym{ura}{URA}{uniform rectangular array}
\newacronym{urllc}{URLLC}{ultra-reliable low-latency communication}
\newacronym{uwb}{UWB}{ultrawideband}
\newacronym{wb}{WB}{wideband}
\newacronym{wpt}{WPT}{wireless power transfer}
\newacronym{zf}{ZF}{zero-forcing}

\newacronym{lpt}{LPT}{laser power transfer}
\newacronym{rfpt}{RFPT}{radio frequency power transfer}
\newacronym{ipt}{IPT}{inductive power transfer}
\newacronym{cpt}{CPT}{capacitive power transfer}
\newacronym{apt}{APT}{acoustic power transfer}
\newacronym{cfo}{CFO}{carrier frequency offset}
\newacronym{lo}{LO}{local oscillator}
\newacronym{if}{IF}{intermediate frequency}
\newacronym{duc}{DUC}{digital up conversion}
\newacronym{ddc}{DDC}{digital down conversion}
\newacronym{dsp}{DSP}{digital signal processing}
\newacronym{vco}{VCO}{voltage-controlled oscillator}
\newacronym{pll}{PLL}{phase-locked loop}
\newacronym{lna}{LNA}{low noise amplifier}
\newacronym{dac}{DAC}{digital-to-analog converter}
\newacronym{adc}{ADC}{analog-to-digital converter}
\newacronym{de}{DE}{drain efficiency}
\newacronym{pae}{PAE}{power-added efficiency}
\newacronym{sar}{SAR}{specific absorption rate}
\newacronym{mppt}{MPPT}{maximum power point tracking}
\newacronym{isi}{ISI}{intersymbol interference}
\newacronym{pps}{PPS}{pulse per second}
\newacronym{icnirp}{ICNIRP}{International Commission on Non-Ionizing Radiation Protection}
\newacronym{fd}{FD}{Front-haul distance}
\newacronym{wd}{WD}{Wireless distance}
\newacronym{ntp}{NTP}{Network Time Protocol}
\newacronym{ptp}{PTP}{Precision Time Protocol}
\newacronym{wr}{WR}{White Rabbit}
\newacronym{wlan}{WLAN}{wireless local-area network}

\begin{document}
\title{Deep Learning for sub-THz Radio Unit Selection using sub-10 GHz Channel Information and Inferred Device Beamforming\vspace{-0.2em}}
\author{Nishant Gupta$^{1,3}$, Muris Sarajlic$^2$ and Erik G. Larsson$^1$ \\
	${}^1$Department of Electrical Engineering (ISY), Link{\"o}ping University, Link{\"o}ping, Sweden \\
    ${}^2$Ericsson Research, Lund, Sweden \\${}^3$Department of Communication and Computer Engineering, LNMIIT Jaipur, India\\
	Email: nishantgupta.nic@gmail.com, muris.sarajlic@ericsson.com, erik.g.larsson@liu.se \vspace{-1.5em}
\thanks{This work has received funding from the EU programmes Horizon Europe (No. 101096302 – 6GTandem). Nishant Gupta is now with Department of Communication and Computer Engineering, LNMIIT Jaipur, India, was with Department of Electrical Engineering (ISY), Link{\"o}ping University, Link{\"o}ping, Sweden, when this work was performed.}}
\maketitle

\begin{abstract}
The dense and distributed deployment of sub-THz radio units (RUs) alongside sub-10 GHz access point (AP) is a promising approach to provide high data rate and  reliable coverage for future 6G applications. However, beam search or RU selection for the sub-THz RUs incurs significant overhead and high power consumption. To address this, we introduce a method that leverages deep learning to infer a suitable sub-THz RU candidate from a set of sub-THz RUs using the sub-10 GHz channel characteristics. A novel aspect of this work is the consideration of inter-band beam configuration (IBBC), defined as the broadside angle between the low-band and high-band antenna patterns of the user equipment (UE). Since IBBC indicates the beamforming information or UE's orientation, it is typically not shared with the network as a part of signalling. Therefore, we propose a solution strategy to infer a suitable sub-THz RU even when UEs do not share their IBBC information. Simulation results illustrate the performance of the inferred sub-THz RU and highlights the detrimental impact of neglecting UE orientation on the systems performance. 
\end{abstract}
\vspace{-0.5em}
\section{Introduction}
The deployment of sub-terahertz (sub-THz) radio units (RUs) is considered crucial for enabling extreme-throughput use-cases in sixth-generation (6G) networks due to its large available bandwidth \cite{9269931}. However, achieving reliable coverage using only sub-THz RUs is challenging due to significant signal attenuation \cite{9887921}. A feasible approach is to overlay the distributed deployment of sub-THz RUs with the traditional sub-10 GHz, where sub-10 GHz provides reliable coverage and sub-THz RUs offers high data-rate links. As a result, the sub-10 GHz overcomes the known drawbacks of sub-THz, ensuring low-rate communication and extending sub-THz coverage \cite{8387211}. This concept is proposed and its practical aspects are investigated in the EU-funded Horizon Europe Research Project 6G Tandem \cite{tandem}.  

Recently, numerous studies have focused on the combined efforts of integrating sub-6GHz and mmWave bands~\cite{8322248, 10292615, 10577648}. In mmWave frequencies, pathloss and low power amplifier (PA) power compromise the link budget. One solution to improve link budget is to ensure a high antenna array gain by employing antenna arrays with a large number of antennas. 
Since the required array gain is high (typically on the order of 20 – 40 dB), the beamwidth is small. This implies that there will be many beams covering an angular coverage sector (typically tens or hundreds of beams). As user equipment (UE) moves, the signal quality in the serving beam may suddenly drop and a new beam needs to be found, typically based on sending reference signals in candidate beams to the UE in the downlink (DL) and the UE providing a measurement report. As in beamsteering typically only one beam at a time can be used for sending reference signals, this implies a large number of scanned beams, and consequently longer beam search time.
In dense and distributed deployment of sub-THz RUs following the radio stripe concept \cite{9499049}, 
 the aforementioned problem is exacerbated by the need to send reference signals in the DL from multiple RUs, and for each RU in multiple beams. Specifically, this leads to high power consumption, as all RUs preceding an RU on the stripe have to be active in order for RU to be able to transmit, and each amplifying RU will consume a non-negligible amount of power 
Moreover, exhaustive search across all the possible beam pairs leads to substantial overhead and extensive latency for the communication systems.

It is therefore of high importance in mmWave systems to narrow down the RU and beam search when needing to update the serving RU in order to reduce signaling overhead, latency, and power consumption. 
Early works in this direction proposed compressed sensing (CS) to reduce the beam search overhead \cite{8792393}. 
In recent years, machine learning (ML) techniques have been proposed to address the beam selection problem \cite{9034044,9048966,9121328,8114345}. For instance, a deep learning algorithm for beam selection in the mmWave band using the sub-6GHz spectrum was proposed in \cite{9121328}. Furthermore, Alrabeiah et al. employed a fully connected neural network for mmWave beam selection, achieving 90\% accuracy in beam selection \cite{8114345}.

However, prior art typically considered the scenarios where the UE is static and modelled the wireless link between the network and the UE without considering beam patterns at the UE. In present generation, the UE's (dual-band UEs) is also equipped with multiple antennas. The antenna pattern reported at the UE may not be shared with the network during the network-device context establishment. Thus, to bridge this gap, in this paper, we present a novel idea to narrow down the search by inferring a suitable sub-THz RU candidates using the lower-band information by exploiting ML where the modelling of the wireless link between the sub-THz and the sub-10 GHz will also depend on the beam patterns at the UE side which accounts for UE's orientation. 
\section{Inferring a suitable RU/Beam}
\begin{figure}[t]
    \centering
    \begin{subfigure}[b]{0.7\linewidth}
        \includegraphics[width=\linewidth]{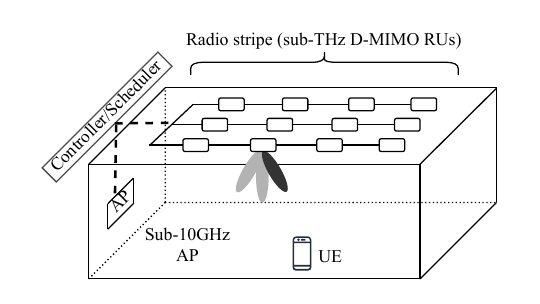}
        \caption{}
        \label{fig:subfig1}
    \end{subfigure}
    \hfill
    \begin{subfigure}[b]{0.25\linewidth}
        \includegraphics[width=\linewidth]{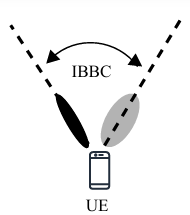}
        \caption{}
        \label{fig:subfig2}
    \end{subfigure}
    \caption{(a) 6G tandem setup comprising of distributed sub-THz RU and sub-10 GHz AP, and (b) concept of IBBC. \vspace{-1em}}
    \label{fig:model_system}
\end{figure}

We consider a dual-frequency (6G tandem) setup that  consists of one sub-10 GHz access point (AP) and distributed sub-THz RUs, each with multiple antenna elements, arranged according to a radio stripe concept, as illustrated in Fig. \ref{fig:model_system}(a). 
Both the sub-10 GHz AP and sub-THz RUs are connected to a central controller or scheduler that manages the exchange of control signals between the two frequency bands. The sub-THz signals are supported by RUs integrated into a dielectric fiber capable of conducting RF propagation. 
The signal to each RUs is generated in the central processing unit and is distributed in the analog form through the fibre to each RU in a multiple stripe setup, thus enabling a sub-THz distributed multiple input multiple output (D-MIMO) system.  Moreover, each RU will either radiate the signal to the UE or amplify the signal and transmit it over the fiber to the next RU. 

The working assumption of the sub-10 GHz frequencies is between 2.5 GHz and 4.2 GHz (referred to as Mid-band in 5G), or any other available frequencies below 10 GHz. Sub-THz frequencies refers to the D-band (110-170 GHz). The sub-THz RU will support short range communications with beam-switching and beam-steering capabilities. The propagation channel between the sub-THz/sub-10 GHz and the UE will depend on the mobility and rotation of the UE, apart from the environment. Thus, the modelling of the wireless link between the network-UE wireless links at ub-THz/sub-10 GHz will also depend on the beam patterns at the UE side. In the following, we first show the general model that can be employed to infer the best RU from the prior art. 

\subsection{sub-THz RU and beam selection assisted  with side information from sub-10 GHz}
\begin{figure}
    \centering
    \includegraphics[width=1\linewidth, trim={0 1.9cm 1cm 0},clip]{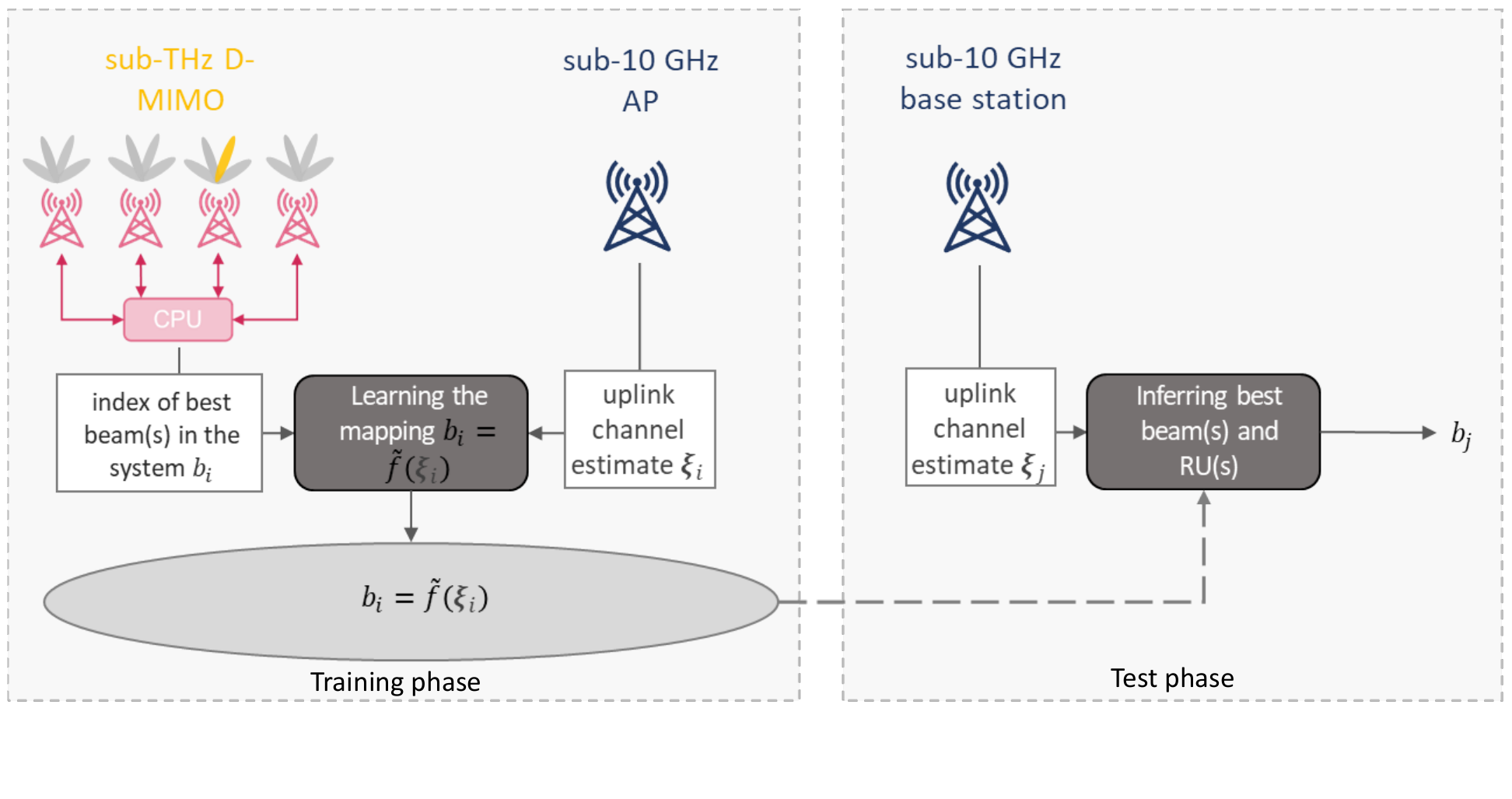}
    \caption{Information flow for inference of best RU from prior art.\vspace{-1em}}
    \label{fig:bestRU_art}
\end{figure}

In a usual implementation, the RU inference at sub-THz would follow these steps: \\
\textbf{Training phase to learn a model} 
    \begin{itemize}
        \item[1)] UE occupying position $i$ sends reference signal in the uplink (UL) of a lower band (sub-10 GHz). Network determines channel property $\zeta_i$ associated with the UL channel (e.g. channel impulse response, angle-delay spectrum, etc.) 
        \item[2)] Simultaneously or nearly simultaneously, the network sends reference signals in the DL from all RUs or a subset of RUs at sub-THz. UE measures signal quality and reports index of best RU $b_i$ or set of $N$ best RUs $b_i= [b_{i,1}, b_{i,2}, \cdots b_{i,N}]$. “Best” here means achieving best signal quality in some sense (e.g. highest received power, highest signal-to-noise ratio, etc.). 
        \item[3)] For a set of positions $i$ (and possibly for different UEs occupying different positions), the network collects pairs $(\zeta_i, b_i)$ and learns an approximation $\Tilde{f}$ of the mapping $b_i = f( \zeta_i )$, typically through a supervised ML approach. 
        \end{itemize}
 \textbf{Using the learned model}
    \begin{itemize}
        \item[1)] A UE, occupying position $j$ sends an UL reference signal in the lower frequency band (sub-10 GHz). Network determines channel property $\zeta_j$. 
         \item[2)] Using the learned mapping $\Tilde{f} ()$, network infers the best RU  $\Tilde{b}_j=\Tilde{f}(\zeta_j)$.
    \end{itemize}
The described procedure has been analyzed in detail in \cite{9034044}, among other works. Fig. \ref{fig:bestRU_art} describes the information flow.

\subsection{Problem with the existing model}

\begin{figure}
    \centering
    \includegraphics[width=1\linewidth]{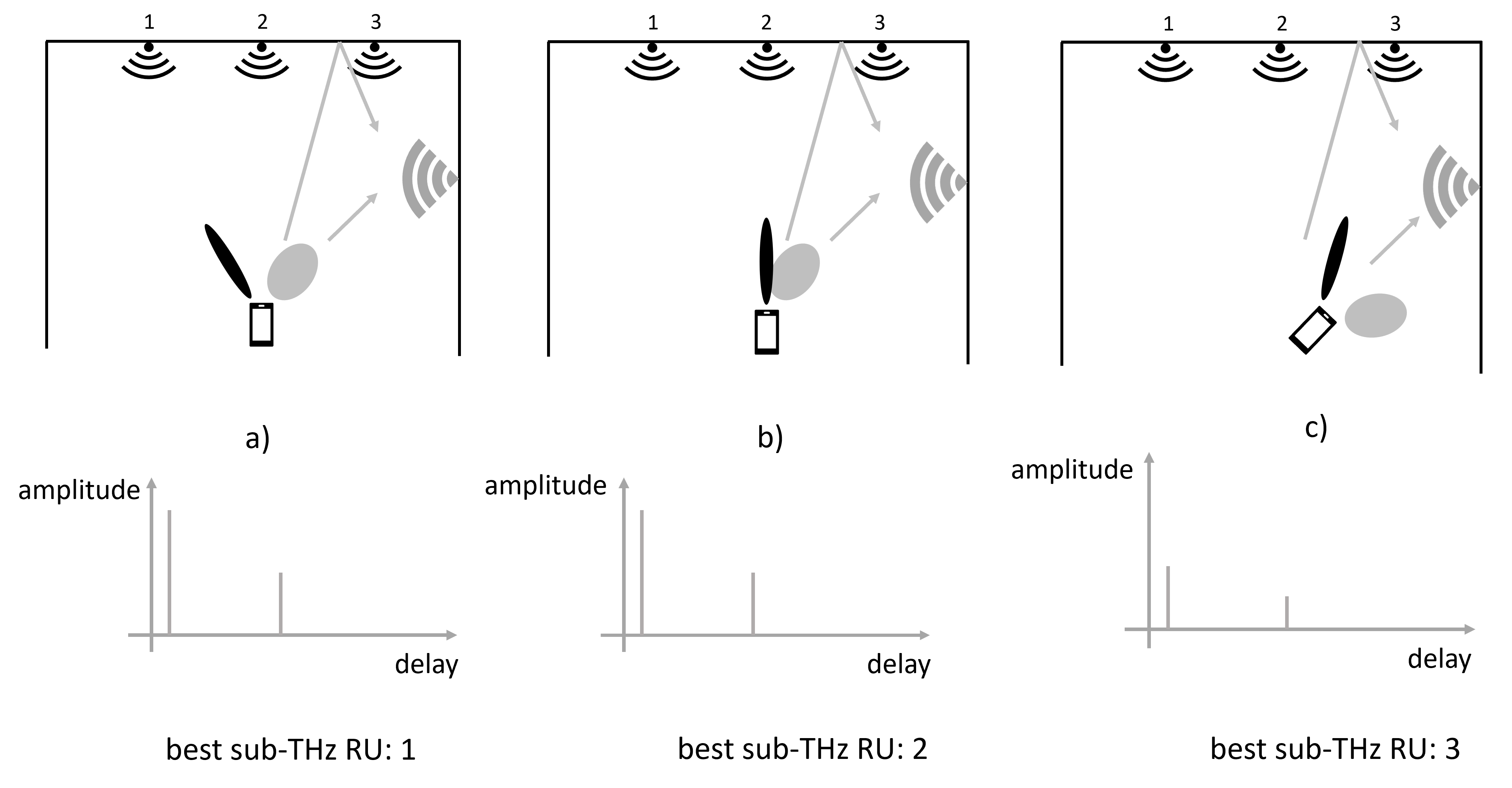}
    \caption{Impact of UE orientation and beam configuration on inference. Top row: a tandem scenario with 3 sub-THz RUs labeled 1-3 and a sub-10 GHz AP. Middle row: impulse responses at sub-10 GHz for each of the scenarios. Bottom row: best RU for each scenario.\vspace{-1em}}
    \label{fig:bestRU_idea}
\end{figure}

The system for inferring best RUs as described in the previous sub-section can provide satisfactory performance with proper choice of ML algorithms and proper training. However, for UEs that use directive antennas and whose orientation randomly rotates in 3D, the approach as described can experience a significant degradation in performance. Fig. \ref{fig:bestRU_idea} helps illustrate the problem. In scenario 1 (Fig.~\ref{fig:bestRU_idea}(a)) the sub-THz beam from UE is illuminating RU-1. The sub-10 GHz beam from UE is pointing towards the sub-10 GHz AP, resulting in an impulse response as seen in the middle row. In scenario 2 (Fig. \ref{fig:bestRU_idea}(b)) the angle between the sub-THz and sub-10 GHz beams at UE has changed, and while the sub-10 GHz beam is unchanged, the sub-THz beam is now illuminating sub-THz RU-2. It can be noted that in Figs.\ref{fig:bestRU_idea}(a) and \ref{fig:bestRU_idea}(b), the same sub-10 GHz impulse response is paired with two different RUs labeled as best, i.e., the function $b_i = f(\zeta_i)$, where the sub-10 GHz impulse response can be used for $\zeta_i$, becomes a one-to-many mapping due to different angles between sub-THz and sub-10 GHz. Further, scenario 3 (Fig.~\ref{fig:bestRU_idea}(c)) keeps the same angle between sub-THz and sub-10 GHz as in Fig.~\ref{fig:bestRU_idea}(a) but the UE is rotated by 45 degrees. Here, although the UE stayed in same position $i$, a different sub-10 GHz channel impulse response $\zeta_i$ is generated, i.e. there is a one-to-many mapping between 3D position of UE and  channel property $\zeta_i$ due to impact of antenna pattern at sub-10 GHz.

The situation complicates further if the beam shape at sub-THz and sub-10 GHz at UE is allowed to change between scenarios (not illustrated in the example). For best performance of the inference scheme as presented in Section II-A, one needs to account for the impact of UE orientation, antenna geometry and antenna pattern on the mapping $b_i = f(\zeta_i)$. Assuming that the radiation patterns of sub-10 GHz and sub-THz do not change between possible scenarios, and defining $\Omega_i$ as the 3D (roll-pitch-yaw) orientation of the device and $\beta_i$ as the 2D (azimuth and elevation) angle between the broadside of the sub-10 GHz and sub-THz antenna radiation patterns (here taking also the array pattern in case of several antennas), 
a more accurate model for the mapping between UE position and configuration and best RU is 
\begin{equation}
    b_i = f(\zeta_i, \Omega_i, \beta_i).
\end{equation}
For the ease of further exposition, we refer to angle 
$\beta_i$ as the inter-band beam configuration (IBBC). Clearly, if the inference is based only on $\zeta_i$ but ignores orientation $\Omega_i$ and IBBC $\beta_i$, the performance of the inference scheme is expected to be poor as ML cannot resolve the multiple one-to-many mappings. 

\textit{Remark:} Note that for a device whose sub-10 GHz and sub-THz beams are fixed (cannot be steered), IBBC $\beta_i$ is determined by the angle between the sub-10 GHz and sub-THz arrays and does not change over time for a particular device design. Sub-THz will, however, typically be steerable over a grid of beams, so IBBC $\beta_i$ may comprise a set of values and change dynamically with time. 

A natural step towards solving problem as outlined above is providing the ML inference algorithm with orientation data $\Omega_i$ and IBBC data $\beta_i$ during model training and model use phases, i.e., making $(\zeta_i, \Omega_i, \beta_i)$ the features that are inputs of the supervised learning algorithm. The key problem here is that some UEs (due to e.g. secrecy) will not be willing to share orientation and especially beamforming information.Additionally, information may be corrupted by estimation noise.  \textit{Summary of problems:} Inference of best RUs needs to take into account UE orientation and beamforming information, and not all UEs can be expected to share such information. 


\section{Proposed Solution}

Key underlying concept in the proposed solution is the concept of IBBC. IBBC can be defined as the angle, in azimuth and elevation, between the broadside of the antenna/array pattern used at low band and high band, at the UE. Fig. \ref{fig:model_system}(b) illustrates the concept of IBBC, presented only in one of the azimuth or elevation. IBBC can be fixed for a particular device (depending on the antenna architecture) or change dynamically as the device chooses different sub-THz beams. As IBBC is essential for the good performance of low-band assisted beam search in sub-THz, the proposed solution aims to estimate IBBC and use it to refine an initial, imprecise inference of the best RU. The focus of the proposed approach are the three ML algorithms implemented at the network side.\\ \textbf{Algorithm 1} takes as one input (feature) the channel characteristic(s) measured between dual-band UE and low-band AP in system. The characteristic can be measured in DL and reported by UE or (preferably) measured in the UL from reference signals sent by the UE. The characteristic can be the channel impulse response, such as power delay profile, angular-delay profile, etc. Another part of the input (feature) is the channel characteristic(s) between one or more high-band RU(s) and the dual-band UE. Algorithm 1 infers the IBBC from the features. Both low and high-band channel characteristics are considered as features as the IBBC is the angle between the beams of low-band and high-band. Thus, these features are required to estimate IBBC. \\ \textbf{Algorithm 2} takes the low-band channel characteristic and IBBC as features, and from these, infers the index of sub-THz RU with the best link quality (measured with SNR).\\ \textbf{Algorithm 3} takes as a feature low-band channel characteristics and infers the index of the best RU or beam. 

Algorithms 1 and 2 form the novel part of this work. Algorithm 3 is known from prior art but is a necessary part of the proposed solution. The proposed method consists of algorithmic steps implemented separately at UE and network.

\begin{figure}[t]
    \centering
    \includegraphics[width=0.6\linewidth]{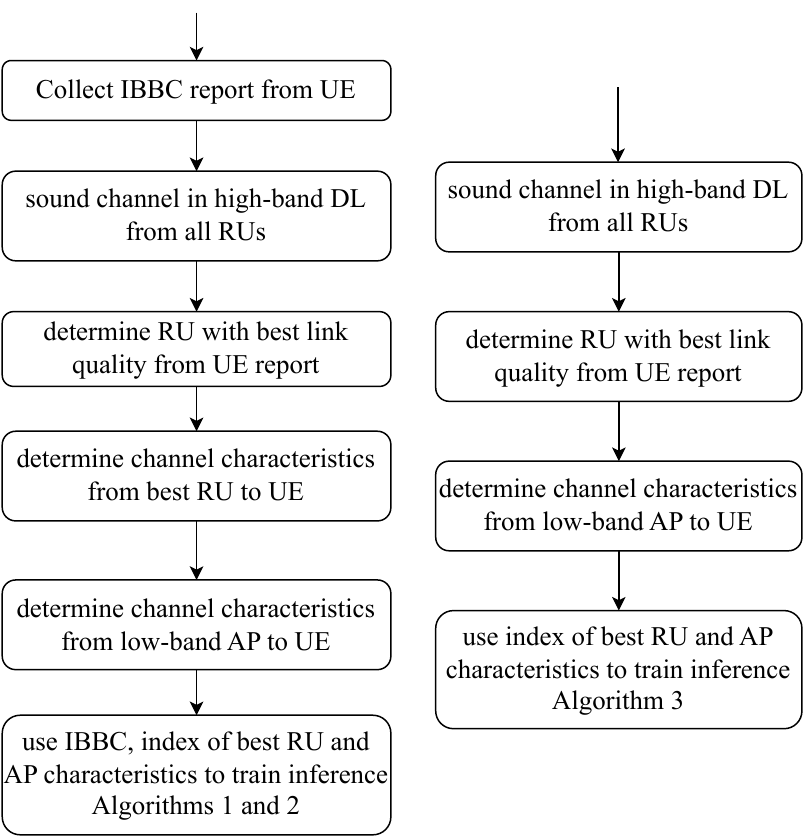}
    \caption{Training of Algorithms 1 and 2 (left-hand side) and 3 (right-hand side).\vspace{-1.5em}}
    \label{fig:bestRU_algorithm}
\end{figure}
\subsubsection{UE aspects}
One novel aspect of this work is the UE informing the network of whether it is capable of providing a report on its currently used IBBC. This capability signaling can form part of the usual capability signaling, e.g. performed during network-device context establishment. For UEs that are capable of providing IBBC, the reporting may be based on a request by the network. Both capability signaling and IBBC reporting require to be implemented as part of the standard. 

\subsubsection{Network Aspects}
As implied by supervised learning, the inference algorithms implemented at network side need to be trained with known labels. For dual-band UEs that can report IBBC, the network collects (receives as report or estimates) IBBC, low-band channel characteristic, and high-band channel characteristic to one or more RUs. In a preferred implementation, the channel characteristic to best RU is estimated. 

The collected data is used to train inference Algorithms 1 and 2. The training can be performed by dedicated UEs that purposefully map out the coverage area, reporting channel characteristics and IBBC as part of the “cell calibration”. Alternatively, the data can be collected gradually, during normal operation, from UEs that happen to have the ability to report IBBC. Additionally,  legacy Algorithm 3 also needs to be trained. Training of Algorithm 3 belongs to prior art but is mentioned here as a supporting embodiment. The steps essential for training of Algorithms 1-3 are given in Fig. \ref{fig:bestRU_algorithm}.

\begin{figure}
    \centering
    \includegraphics[width=0.8\linewidth, trim={0 0.3cm 0cm 0},clip]{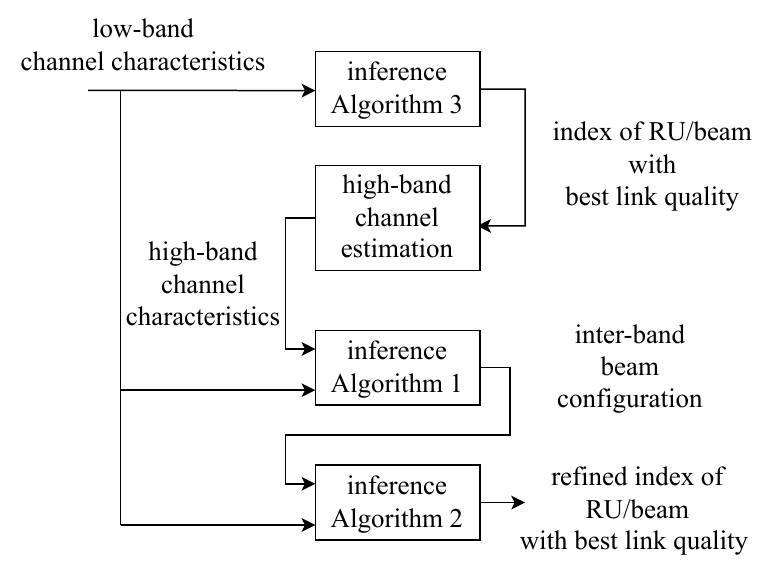}
    \caption{Steps and information flow of the proposed 3-step approach. \vspace{-1em}}
    \label{fig:bestRU_algorithm3}
\end{figure}

The novel part of this work is a 3-step method, implemented at the network side, of refining the estimate of best RU. For the UE that can report IBBC, we can directly employ Algorithm 2 to obtain the best RU. However, the problem arise when the UE cannot report IBBC. This method is performed for UEs that cannot report IBBC. The method is illustrated in Fig. \ref{fig:bestRU_algorithm3} and consists of following steps:  
(i) Performing coarse inference of the best RU from the low-band channel characteristic using inference Algorithm 3. Then, estimating the channel characteristic between the best RU and the dual-band UE. The low-band connection can be used for the report as it is typically more reliable; (ii) To infer IBBC of the dual-band UE using the low-band and high-band channel characteristics (as estimated in step 1), inference Algorithm 1 is used. (iii) To infer the index of the best RU from the low-band channel characteristic and the inferred IBBC, inference Algorithm 2 is used. The newly inferred RU may or may not be the same as the one inferred in step 1. With the inference being based on more information, the aim is to have a more precise inference of the best RU and thus better performance. 
The method is applied at a UE and at the network side, and the goal is to infer the best sub-THz RU for serving the UE in the downlink, based on the channel properties in the UL of a sub-10 GHz system, and on additional information. 

By refining the inference of the best RU based on the inferred IBBC, the performance of the best RU inference algorithm is improved for UEs that cannot report IBBC. IBBC can be understood as a representation of the UE beamforming configuration (choice of sub-THz beam), which UEs typically choose not to report. For those UEs, the low-band assisted RU finding scheme will not perform well. The proposed solution can thus be considered a key enabler of the low-band assisted sub-THz paradigm for UEs that apply beamforming (which will be a vast majority of UEs at sub-THz). 

\section{Performance Evaluation}
The simulation setup consists of one sub-10 GHz AP and nine sub-THz RUs deployed in an office environment of dimension (8 x 5 x 3) meters (m) in length, width, and height, respectively. The sub-10 GHz AP is positioned at the center of an 8 m wall, and the RUs are placed at the ceiling. The UEs are distributed throughout the room with a 0.5 m separation between them, and their height is restricted to 1.5 m. The transmit antennas at both the sub-10 GHz and sub-THz are a 4-element uniform rectangular array (URA) with the specific patterns. The receiving antenna at UE is also 4-element URA. The transmitter frequencies are 5.8 GHz for sub-10 GHz and 100 GHz for sub-THz. We consider a total of 16 IBBCs with different UEs reporting one of the 16 different possible IBBCs to the network. In order to expand the size of the training dataset, each UE position on the (0.5 x 0.5) m grid carries four IBBC, randomly chosen from the set of 16 IBBCs. The multipath channel model between the transmitter and receiver is built using ray tracing. The channel characteristics in form of power delay profile is measured between the dual-band UE, low-band AP and high-band RU. At each position, we compute four pairs of $b_i = f(\zeta_i,\Omega_i,\beta_i)$, corresponding to different $\beta_i$, which are used to train deep learning model. \textcolor{black}{The generated data set is divided into training and validation set, with an 80\% and 20\% ratio, respectively.}
\begin{figure}
    \centering
    \includegraphics[width=0.95\linewidth]{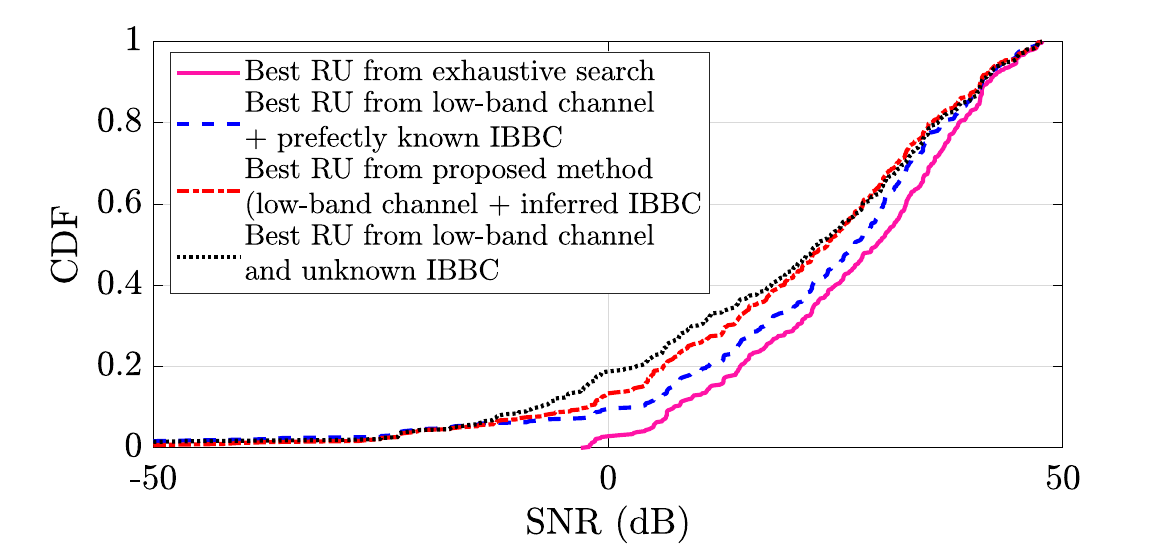}
    \caption{Performance evaluation of the proposed method. \vspace{-1em}}
    \label{fig:perf1}
\end{figure}

\textcolor{black}{The proposed DNN model utilizes three inference classifiers (Algorithms 1-3).} Each classifier comprises 5 layers, including the feature input layer and the classification output layer. The model consists of three hidden layers containing 128, 50, and 8 neurons, respectively. The activation function used in the hidden layers is the Rectified Linear Unit (ReLU). \textcolor{black}{The learning rate is 0.01, and the optimizer employed is Stochastic Gradient Descent with Momentum (SGDM).} The size of the input layer is equal to (i) the combined size of the low-band channel characteristics and high-band channel characteristics for Algorithm 1, (ii) the combined size of the low-band channel characteristics and UE's IBBC information for Algorithm 2, and (iii) the size of the low-band channel characteristics for Algorithm 3.  Similarly, the size of the output layer is equal to (i) size of the IBBC (azimuth and elevation angles) for Algorithm 1, and (ii) the index of the best RU for Algorithms 2 and 3. Moreover, the classification output layer employs a softmax function as its activation function. The channel characteristic of sub-10 GHz channel chosen as input feature is the instantaneous PDP, i.e. magnitude squared of the impulse response. That is, inference of best beam at sub-THz is based only on delay information from sub-10 GHz, without taking into account angular information as in \cite{9034044}. 

Performance evaluation results are shown in Fig. \ref{fig:perf1} as CDF of the SNR at the UE (receiver). It can be observed from Fig. \ref{fig:perf1} that the proposed method (dashed-dotted curve) manages to partially close the performance gap between Algorithm 3 (i.e. using RU inferred as best from low-band channel and with unknown IBBC), represented by the dotted curve, using the RU inferred from low-band channel and perfectly known IBBC (dashed curve). Performance improvement over Algorithm 3 at lower percentiles is 4-5 dB. 

\begin{figure}
    \centering
    \includegraphics[width=0.8\linewidth]{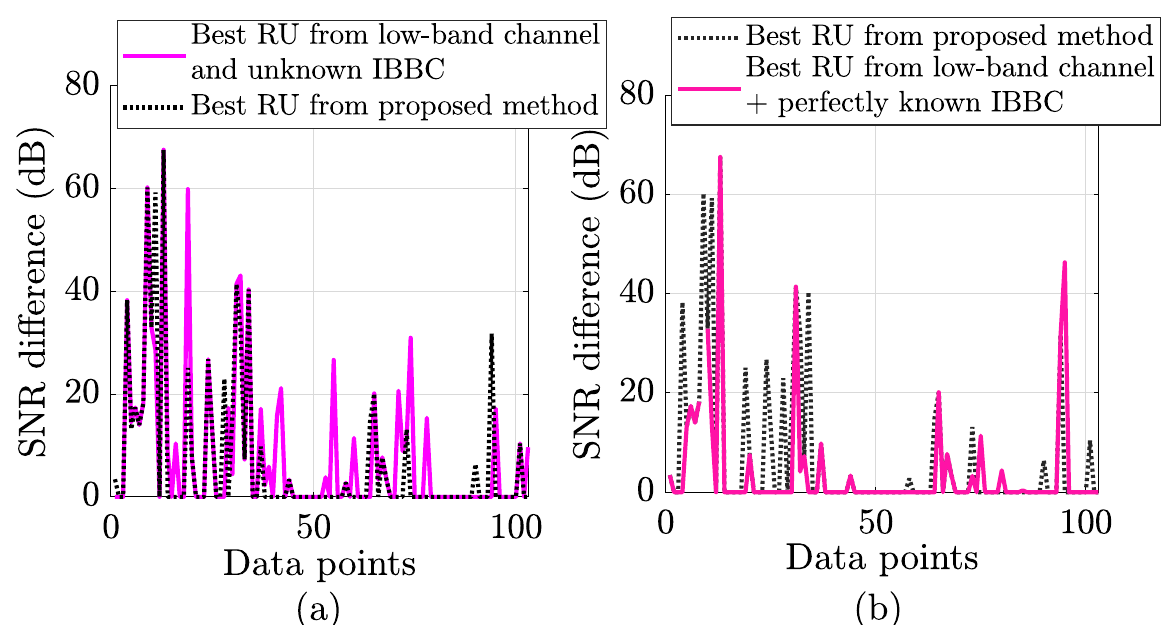}\vspace{-0.3em}
    \caption{SNR difference of proposed method with different scenarios.}
    \label{fig:difference}
\end{figure}

Fig. \ref{fig:difference} illustrates the difference in SNRs at each data point for various approaches. It can be observed from Fig. \ref{fig:difference}(a) that the SNR difference reduces using the proposed approach where we first infer the IBBC to predict the best RU compared to scenarios where IBBC information is not considered. Similarly, Fig.  \ref{fig:difference}(b) demonstrates that using the fully available IBBC for network training significantly reduces the SNR difference compared to the proposed method. However, the SNR differences in Fig. \ref{fig:difference}(b) are relatively small, indicating the effectiveness of the proposed approach and the advantages of incorporating IBBC information. 

\begin{figure}[t]
    \centering
    \includegraphics[width=0.9\linewidth]{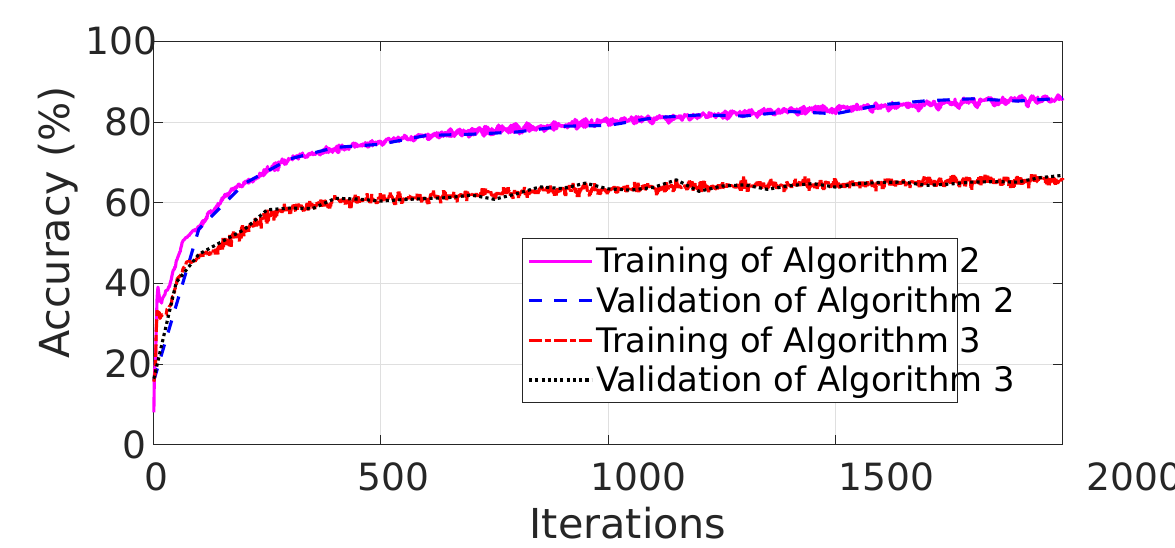}
    \caption{Training accuracy of Algorithm 2 and Algorithm 3, i.e., the inference of the best RU.}
    \label{fig:val1}
\end{figure}

\begin{figure}[t]
    \centering
    \includegraphics[width=0.87\linewidth]{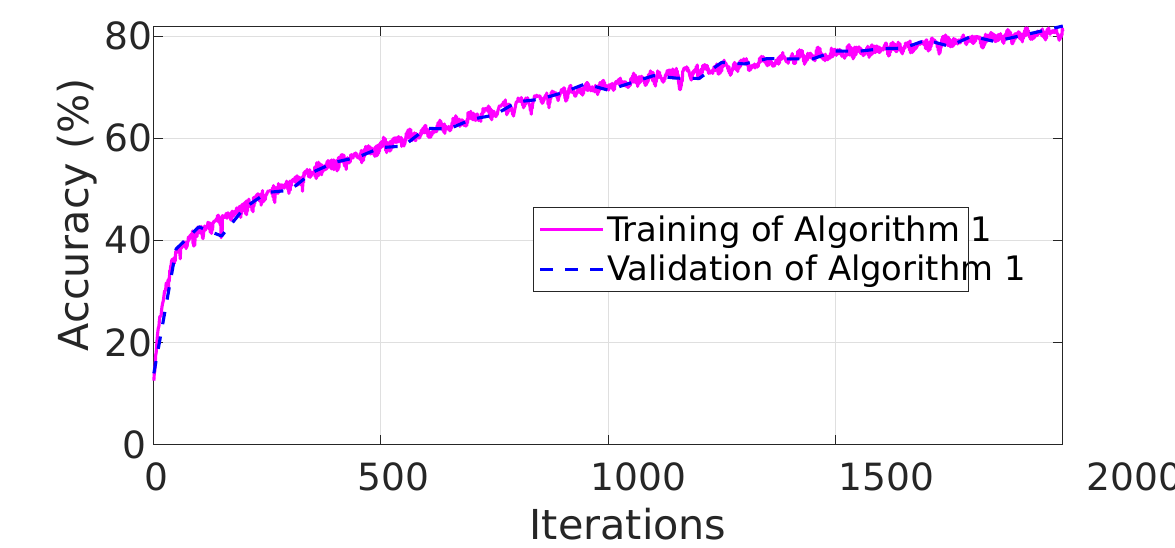}
    \caption{Training accuracy of Algorithm 1 i.e., inferred IBBC based on the low-band and high-band channel characteristics.\vspace{-1em}}
    \label{fig:val2}
\end{figure}

To assess the performance of the trained deep learning algorithms, we plot Figs. \ref{fig:val1} and \ref{fig:val2}. Fig. \ref{fig:val1} depicts the accuracy of Algorithm 2 and Algorithm 3. Notably, Algorithm 2 shows higher accuracy compared to Algorithm 3, which relies solely on low-band characteristics. This improvement is due to the fact that, for a dual-band UE, the sub-THz beams may point in different directions at different times, resulting in one-to-many mapping in case of Algorithm 3 (as illustrated in Fig. \ref{fig:bestRU_idea}) while the sub-10 GHz beams point consistently. Thus showcasing the impact of the IBBC on the system's performance and the trained network using the IBBC yields a more accurate result. 
Finally, Fig. \ref{fig:val2} shows the accuracy of Algorithm 1. 


\section{Conclusion}
In this paper, we proposed a deep learning-based method for selecting the optimal sub-THz RU using the sub-10 GHz channel information to ensure high SNR for UE. A key underlying concept in the proposed solution was IBBC (which captures the effects of beamforming at the UE side), defined as the angle between the broadside of the antenna pattern used at low-band and high band at the UE. While some UEs may share the IBBC information, however, the problem arises when the IBBC information is not part of the signalling. Our proposed solution strategy, wherein IBBC is inferred based on a trained separate model, infers the index of the best sub-THz RU even without shared IBBC information. Simulation results demonstrate that inferring the best RU with the use of inferred IBBC as side information helps improve the performance compared to the case where IBBC is not considered. For fast-moving scenarios, frequent channel estimates are necessary to counteract rapid channel variations.

\section*{Acknowlegement}
This work has received funding from the EU programmes Horizon Europe (No. 101096302 – 6GTandem). Nishant Gupta is now with Department of Communication and Computer Engineering, LNMIIT Jaipur, India, was with Department of Electrical Engineering (ISY), Link{\"o}ping University, Link{\"o}ping, Sweden, when this work was performed.

\bibliographystyle{IEEEtran}
\bibliography{ref}

\end{document}